\documentclass[fleqn,usenatbib,useAMS]{mnras}
\pdfoutput=1
\usepackage{aas_macros,bm,amsfonts}
\usepackage{graphicx,ulem}
\usepackage{rotating}
\usepackage{float}
\usepackage{etoolbox}
\usepackage{hyperref,epstopdf}

\def \epm {\varepsilon_{\rm M}}
\def \II {\mathbb{I}}

\def \dl {d\bm{\ell}}
\def \Vrec {V_{\rm rec}}
\def \Np {N_{\rm p}}
\def \VA {V_{\rm A}}

\def \ff {{\bm f}}
\newcommand{\eee}{\hat{\bm e}}
\def \fh {\tilde{{\bm f}}}

\def \XX {{\bm X}}
\def \VV {{\bm V}}
\def \Xdot {\dot{{\bm X}}}
\def \Vdot {\dot{{\bm V}}}
\def \xx {{\bm x}}
\def \kk {{\bm k}}
\def \ii {{\rm i}}

\def \del2z {\partial^{2}_{z}}

\def \AAA {{\bm A}}

\def \UU {{\bm U}}

\def \JJ {{\bm J}}
\def \Lx {L_{\rm x}}
\def \Ly {L_{\rm y}}
\def \Lz {L_{\rm z}}
\def \Lzbytwo {\frac{L_{\rm z}}{2}}
\def \gmax {\gamma_{\rm max}}
\newcommand{\SSSS}{\mbox{\boldmath ${\sf S}$} {}}

\def \grad {{\bm \nabla}}
\def \curl {{\bm \nabla} \times}

\def \delt {\partial_t}
\def \Dt {D_t}

\newcommand{\eq}[1]{(\ref{#1})}

\newcommand{\avgz}[1]{\left\langle #1\right\rangle_{z}}

{}

\def \Rm  {\mbox{Re}_{\rm M}}

\def \Rey  {\mbox{Re}}
\def \S  {\mbox{S}}

\def \kf  {k_{\rm f}}

\def \urms  {u_{\rm rms}}

\def \BB {\bm B}
\def \EE {\bm E}

\def \curl {{\bm \nabla}\times}

\def \Epsilon {\epsilon_{\rm L}}

\def \Rm  {\mbox{Re}_{\rm M}}
\def \Lo  {\mbox{Lo}}
\def \Ma  {\mbox{Ma}}
\def \VA  {V_{\rm A}}
\def \omegac  {\omega_{\rm c}}
\def \MaA  {\mbox{Ma}_{\rm A}}
\def \Pm  {\mbox{Pr}_{\rm M}}
\def \cs  {c_{\rm s}}
\def \kf  {k_{\rm f}}

\def \tauL {\tau_{\rm L}}

\def \urms {u_{\rm rms}}

\def \BB {\bm B}

\def \cP {\mathcal{P}}
\def \cQ {\mathcal{Q}}

\def\onethird{{\textstyle{1\over3}}}

\def \half {{\textstyle{1\over2}}}
\def \Aini {\bm{A}^{\rm ini}}
\def \Bini {\bm{B}^{\rm ini}}
\def \Biniy { B^{\rm ini}_y}
\def \Azini { A^{\rm ini}_z}
\newcommand{\fig}[1]{Fig.~\ref{#1}}

\def\drawing #1 #2 #3 {
\begin{center}
\setlength{\unitlength}{1mm}
\begin{picture}(#1,#2)(0,0)
\put(0,0){\framebox(#1,#2){#3}}
\end{picture}
\end{center} }

\def \apj {ApJ}
\def \apjl {ApJL}

\newtoggle{psfig}
\togglefalse{psfig}

\begin{document}
\title[Energisation by Fast Reconnection]{On the energisation of charged particles by fast magnetic reconnection}
\author[Sharma, Mitra \& Oberoi]{Rohit Sharma$^1$\thanks{E-mail: rohit@ncra.tifr.res.in},
Dhrubaditya Mitra$^2$, Divya Oberoi$^1$\\
$^1$National Centre for Radio Astrophysics, Tata Institute of Fundamental Research, Pune 411 007, India\\
$^2$Nordita, KTH Royal Institute of Technology and Stockholm University,
Roslagstullsbacken 23, SE-10691 Stockholm, Sweden\\
}

\maketitle
\begin{abstract}
We study the role of turbulence in magnetic reconnection, within the
framework of magneto-hydrodynamics, using three-dimensional direct
numerical simulations. For small turbulent intensity we find that the
reconnection rate obeys Sweet-Parker scaling. For large enough
turbulent intensity, reconnection rate departs significantly from
Sweet-Parker behaviour, becomes almost a constant as a function of the
Lundquist number. We further study energisation of test-particles
in the same setup. We find that the speed of the energised particles obeys
a Maxwellian distribution, whose variance also obeys Sweet-Parker
scaling for small turbulent intensity but depends weakly on the
Lundquist number for large turbulent intensity. Furthermore, the
variance is found to increase
with the strength of the reconnecting magnetic field. 
\end{abstract}
\section{Introduction} 

Energetic charged particles with a wide range of energies are
ubiquitous in astrophysics. At one end of the energy scale lie the
galactic cosmic radiation with energies of 
the order of $10^{20}$eV;
at the other end lie particles, 
with energies of the order of  a few  $KeV$s, 
accelerated by the magnetosphere of the Earth.
The energy of particles energised by the Sun lies in the intermediate range:
from particles energised by flare-associated events with energies of
the order of $10^{10}$eV, down to barely detectable events with energies about 
$1$MeV~\citep[see e.g.,][for a review]{fic+mcd66,sweet69,hud+rya95}. 
The production of energetic particles in solar processes seems to
occur  as discrete events \citep[see, e.g.,][]{kli15}.
The current state-of-the-art of the instrumentation spanning observing
bands all the way from $\gamma$-rays to low radio frequencies and
emission mechanisms from thermal bremsstrahlung to coherent plasma 
emissions, allows us to study this process, from the development of 
active regions to the occurrence of the flare itself and the following
aftermath in unprecedented detail. 
We have {\em in-situ} measurements of the plasma properties, usually
at a few isolated points in the vicinity of the Earth.
Nevertheless, the mechanism of energisation for the galactic cosmic
rays, the solar particles or even the magnetospheric particles are
not well-understood.

Theoretically, magnetic reconnection is considered to be one
of the promising mechanisms to energise charged particles; indeed a significant
amount of research in reconnection is motivated by the study of energetic
particles in solar flares.
The phenomenon of magnetic reconnection is one of the fundamental 
processes in astrophysics,~\cite[see e.g.,][and references therein.]{zwe+yam09}
and is worth studying in its own right. 
Magnetic reconnection is typically defined as a process
which gives rise to 
``\ldots a topological rearrangement of the magnetic field that
converts magnetic energy to plasma energy.''~\citep{zwe+yam09}.
Within the domain of applicability of this definition, magnetic
reconnection can be of various varieties.
In the simplest case of resistive (or collisional) reconnection, which 
is applicable for very small but non-zero magnetic diffusivity ($\eta$), 
the magnetic field lines are {\it frozen} in the flow except when 
they reconnect through magnetic diffusivity. 
This is modelled within the magnetohydrodynamic (MHD) description of 
the plasma. 
In the {\it collisionless}, non-MHD, regime, one realises that the
electrons and the ions can have very different time scales and
consequently a two-fluid description is necessary. 

In what follows, we limit ourselves to the MHD description of reconnection.  
Within a reconnection model, if the reconnection rate goes to zero as
$\S \to \infty$ then the reconnection process is defined to be {\it slow};
otherwise, it is called {\it fast}. 
Here $\S \equiv \VA\delta/\eta$ is the Lundquist number with $\delta$
the thickness of  the reconnecting current sheet and $\VA$ the Alfven
speed  given by the strength of the reconnecting magnetic field. 
Clearly, slow reconnection is not of relevance in most astrophysical problems. 
The original quasi-stationary  model for  resistive reconnection by \cite{swe58} 
was shown by \cite{par73} to be slow -- the reconnection rate 
$\gamma \sim \S^{-1/2}$.
Approximate analytical theory by \cite{pet64}, who changed the planar
the geometry of the Sweet-Parker model to that of an ``X'',  does give  
fast reconnection -- the maximum reconnection rate $\gamma \sim 1/\ln \S$ -- 
although  numerical simulations generally show such configuration to be 
unstable unless $\eta$ is not a constant but increases near the 
X-point~\citep{zwe+yam09}. 
Through a boundary layer calculation in a similar geometry, 
\cite{mof+hun02} showed that the evolution of the magnetic flux
-- the rate-of-change of magnetic flux is a measure of the reconnection rate -- 
is determined by two timescales, not just the diffusive time scale but also the  
strain-rate of the flow at the X-point.  
In any case, all the evidence, both numerical and analytical, support the
statement that resistive magnetic reconnection is a slow process unless
small scale turbulence is taken into account.

In two-dimensions (2D), a numerical attempt to study reconnection rate 
as a function of Lundquist number for different turbulent intensities
was made by \cite{lou+uzd+sch+cow+you09}, who showed that at large enough 
turbulent intensities reconnection can be fast -- the reconnection rate becomes 
independent of the Lundquist number. It is believed that such fast reconnection
is due to 2D plasmoid instability~\citep{lou+sch+cow07,hua+bha13}. 
As 2D MHD can be quite different from its three-dimensional (3D) counterpart
a different mechanism may be responsible for fast reconnection in 3D.
It has been argued that reconnection in the presence
of turbulence, in 3D, would be typically fast~\citep{laz+vis99,eyi+laz+vis11}
and has been found to be so in numerical simulations
of both forced~\citep{kow+laz+vis+otm09} and self-generated~\citep{ois+mac+col+tam15} turbulence. 

Through reconnection the magnetic energy is dissipated to energise the ions and electrons.
The crucial question here is whether energisation by reconnection can generate
{\it suprathermal} charged particles, i.e., those with energies ``very
much in excess of their general thermal background''~\citep{par58b}.  
An important class of solar active emission at low radio frequencies
is believed to arise from coherent plasma emission mechanisms.  These
emissions require the presence of a suprathermal population 
of particles in a thermal background, and it is the particles in 
this so called bump-on-the-tail part of the distribution which is responsible 
for the electromagnetic radiation.
The principal question that we want to understand is, whether the process of 
fast magnetic reconnection can produce a family of energized charged particles  
with such properties. 

This is a difficult question to study numerically because in-principle
one has to also take into account the electromagnetic fields created
by the energised ions and electrons, which can be done by performing
kinetic simulations, 
e.g., by using particles-in-cell (PIC) 
algorithms~\citep[see, e.g.,][]{zen+hos07,hos12,hos+muk+ter+shi01,
sir+spi14,guo+li+dau+liu14}.
Another option is to work with the {\it test-particle}
approximation~\citep[see,
e.g.,][]{amb+mat+gol+pla88,kow+dal+laz11,kowal2012particle,deg+kow15,
del+dal+kow16}
where one solves numerically the partial differential 
equations of MHD and uses the resultant electromagnetic field to calculate the
energisation of a number of charged particles, ignoring the electromagnetic
field generated by these energized charged particles. 
Furthermore, an additional approximation -- quasi-stationary --  is
also often employed, where one assumes 
that the evolution of the particles are so fast that time-evolution of
the fields (velocity and magnetic) can be ignored or 
approximated~\citep[see, e.g.,][for a recent example]{thr+bou+neu+par16}.

The rest of this paper is organised as follows: 
in section \ref{sec:model} we describe our reconnection simulation 
which uses the tearing-mode setup of \cite{lou+uzd+sch+cow+you09} but in 
3D. 
Next, we describe how we incorporate test-particles in our setup.
Note that we do not use the quasi-stationary approximation. 
In section \ref{sec:results} we 
measure the reconnection rate by calculating the rate-of-change of 
magnetic flux and then study the 
energisation of the test-particles by calculating the probability distribution 
function (PDF) 
of their speed when they reach the boundary of our simulations box.
We also find out how this PDF depend on various parameters of our model, 
including, the Lundquist number, the intensity of turbulence and the 
strength of the reconnecting magnetic field.  

\section{Model}
\label{sec:model}
Let us first describe our setup without the test-particles. 
We solve the equations of isothermal MHD for the velocity $\UU$,
the magnetic vector potential $\AAA$, and the density $\rho$,
\begin{equation}
\rho\Dt\UU=\JJ\times\BB-\cs^2\grad\rho+\grad\cdot(2\nu\rho\SSSS) +\rho\ff,
\label{eq:mom}
\end{equation}
\begin{equation}
\delt\rho=-\grad\cdot\rho\UU,
\label{eq:cont}
\end{equation}
\begin{equation}
\delt\AAA=\UU\times\BB+\eta\nabla^2(\AAA - \Aini),
\label{eq:ind}
\end{equation}
where the operator $\Dt \equiv \delt+\UU\cdot\grad$ denotes
the convective derivative,
$\BB=\curl\AAA$ is the magnetic field,
$\JJ=\curl\BB/\mu_0$ the current density,
$\mu_0$ is the permittivity of vacuum,
${\sf S}_{ij}=\half(U_{i,j}+U_{j,i})-\onethird\delta_{ij}\grad\cdot\UU$
is the traceless rate-of-strain tensor (the commas denote
partial differentiation),
$\nu$ the kinematic viscosity,
$\eta$ the magnetic diffusivity,
and $\cs$ the isothermal sound speed.
In addition, we assume the ideal gas law to hold.
Our domain is a Cartesian box of size $\Lx\times\Ly\times\Lz$
with $\Lx=\Ly=\Lz=L=2\pi$.
The term $\nabla^2 \Aini$ in \eq{eq:ind} is employed in some but not all of our simulations. 
The aim is to preserve the initial configuration of the magnetic field, such 
that a statistically stationary state of repeated reconnections can be set up,
such that meaningful statistical averages can be computed. 

\subsection{Forced turbulence}
Turbulence is generated by the last term in \eq{eq:mom}, 
with the external force $\ff$ given by \citep{B01}
\begin{equation}
\ff(\xx,t)=\,{\rm Re}\{N\fh(\kk,t)\exp[\ii\kk\cdot\xx+\ii\phi]\},
\end{equation}
where $\xx$ is the position vector.
On dimensional grounds, we choose
$N=f_{0} \sqrt{\cs^3 |\kk|}$, where $f_0$ is a
non-dimensional forcing amplitude which controls the 
intensity of turbulence .
At each timestep we select randomly the phase
$-\pi<\phi\le\pi$ and the wavevector $\kk$
from many possible wavevectors
in a certain range around a given forcing wavenumber, $\kf$.
Hence $\ff(t)$ is a stochastic process that is white-in-time. 
The Fourier amplitudes,
\begin{equation}
\fh({\kk})=\II\cdot\fh({\kk})^{\rm(nohel)} \/,
\end{equation}
where $\II$ is the identity matrix,
and
\begin{equation}
\fh({\kk})^{\rm(nohel)}=
\left(\kk\times\eee\right)/\sqrt{\kk^2-(\kk\cdot\eee)^2},
\label{nohel_forcing}
\end{equation}
is a non-helical forcing function, and $\eee$ is an arbitrary unit vector
not aligned with $\kk$ and $\hat{\kk}$ is the unit vector along $\kk$; note that $|\fh|^2=1$.

\subsection{Test-particle approximation}
The test-particles satisfy the non-relativistic dynamical equations:
\begin{eqnarray}
 \Xdot &=& \VV \label{eq:xdot} \\
 \Vdot &=& \frac{q}{m}\left[ \EE(\xx)\delta^3(\XX-\xx) + \VV\times\BB(\xx)\delta^3(\XX-\xx)\right] \label{eq:vdot}
\end{eqnarray}
where $\XX$ and $\VV$ are the position and velocities of the test particles, which has charge $q$ and
mass $m$, $\delta^3(\cdot)$ is the three-dimensional Dirac delta function. 
The electric field    
\begin{equation}
\EE = -\left[ \UU\times\BB - \eta\JJ \right]
\end{equation}
and the magnetic field $\BB$ are functions of the space coordinate $\xx$. 

\subsection{Initial and boundary conditions}
\label{inbc}
We impose periodic boundary condition on all three directions on all the field 
variables (velocity, density and magnetic field). 
We start our simulations with $\UU=0$ everywhere and with constant
density, $\rho_0$. 
In order to set-up initial magnetic reconnection configuration, we
choose tearing mode configuration, i.e., the initial value for the
vector potential is set as 
$\Aini=(0,0,\Azini)$ with  
\begin{equation}
\Azini = A_0\frac{1}{\cosh^{2}(x/\delta)}
\label{eq:tear}
\end{equation}
where $\delta$ is the width of the current sheet. 
This implies that as a initial condition, only the $y$ component of the
magnetic field is non-zero and is a function of $x$ alone, i.e., 
$\Bini=(0,\Biniy(x),0)$. 
Once the simulation has reached a statistically stationary state of
repeated reconnection events, we introduce the test-particles at
uniformly distributed random locations within the current 
sheet (of thickness $\delta$) with zero initial velocity. 
When a test particle reaches the
boundary of the domain it is removed from the simulation. 
This numerical experiment is repeated several times to obtain better statistics. 
\subsection{Nondimensional parameters}
The usual process of non-dimensionalization of the dynamic equations yield the following
non-dimensional parameters, 
Mach number, $\Ma \equiv \urms/\cs$, 
Alfvenic Mach number $\MaA \equiv \urms/\VA$, 
Lundquist number $\S \equiv \VA\delta/\eta$,
Reynolds number, $\Rey \equiv \urms/(\nu\kf)$,
magnetic Reynolds number, $\Rm \equiv \urms/(\eta\kf)$, 
and the magnetic Prandtl number, $\Pm \equiv \nu/\eta$.
Here we have defined $\urms$ as the root-mean-square velocity of the
flow, and the  $\VA \equiv B_0/\sqrt{\rho_{\rm 0}\mu_{\rm 0}}$,
the Alfven speed, where $B_0$ is the maximum value of the magnitude 
of the reconnecting magnetic field, 
We also obtain two ratios of length scales $\Epsilon \equiv L\kf$
where $L$ is the length of our box which is equal to the length of the
current sheet, and $\epsilon \equiv \kf\delta$. 
We non-dimensionalize the current by $B_0/\delta$.

 The non-dimensionalization of the 
test-particle equations gives one more dimensionless parameters, 
the Lorentz number, $\Lo \equiv \tauL\omegac$
where the cyclotron frequency $\omegac \equiv qB_0/m$ and 
the large-eddy-turnover-time $\tauL \equiv 1/(\kf\urms)$.

Not all the dimensionless numbers listed above are independent of
one another, e.g., $\Pm = \Rey/\Rm$ and $\S = \Rm/\MaA $
In all the runs in this paper, the parameters $\kf,\Epsilon$
and  $\epsilon$ are kept fixed as $3, 6\pi$ and $3$ respectively. 
The numerical values of all the parameters used in our runs are 
listed in table~\ref{tab:para}.
The simulations are run with $N^3$ grid points. To check 
numerical convergence, we have used $N=128$ and $256$ for
all our runs, and $512$ for a few selected runs. 
Note that, although the Reynolds number we use is moderate 
(less than $\sim 160$) our simulations cover a 
large range -- two decades -- of magnetic Reynolds number.

\subsection{Numerical implementation}
We solve the dynamical equations using the pencil-code,
\url{http://pencil-code.nordita.org}, see also \cite{bra+dob02},
 which uses sixth-order central finite-difference in space and
 third-order  Williamson-Runge-Kutta scheme~\citep{wil80} in time. 
The external force  $\ff(t)$ is a white-in-time stochastic process
integrated by using the Euler--Marayuma scheme \citep{hig01}.
We use a uniform grid of $N^3$ points and $\Np=320,000$ test-particles per run.
To solve for each test-particle we need to know $\UU, \EE$ and $\BB$ at the typically off-grid position
of the particles; this is obtained by trilinear interpolation~\footnote{It has been noted by \cite{amb+mat+gol+pla88}
that a cubic-spline interpolation and a trilinear interpolation give same results in a similar
problem.} from neighbouring grid points.  
\begin{table}
\caption{ \label{tab:para}
List of parameters for all the runs. The runs with prefix {\tt A} have the 
lowest turbulent intensity and those with prefix {\tt C} the highest. 
Within each family({\tt A}, {\tt B} or {\tt C}), 
the magnetic Reynolds number, and the Lundquist number increases from
with the suffix {\tt 1} to {\tt 5}. 
For the {\tt A} family, the maximum reconnection
rate, $\gmax$,  shows Sweet-Parker scaling as a function of $\S$.  
The simulations are run with $N^3$ number of grid points with $N=128$ and $256$;
and $\Np=320,000$ number of test-particles for each run.  
The parameters $\Epsilon=2\pi\times3$, and $\epsilon=3$ are kept fixed. 
}\vspace{12pt}
\centerline{\begin{tabular}{ccccccccc}
Run & $f_0$& $\gmax$ & $\Rey$ &$\Rm$ & $\Ma$ & $\MaA$ & $\Pm$   & Lo \\
\hline
\hline
{\tt A1}  &  0.001  &   9.5$\times$10$^{-3}$ &  83 & 83  &  0.067  &  0.65 & 1 & 9 \\
{\tt A2}  &  0.001  &   6.1$\times$10$^{-3}$  &  43 & 144  &  0.034  &  0.34 & 3 & 18 \\
{\tt A3}  &  0.001  &   4.5$\times$10$^{-3}$  &  15 & 158  &  0.013  &  0.12 & 10 & 51 \\
{\tt A4}  &  0.001  &   3.1$\times$10$^{-3}$  &  7 & 261  &  0.006  &  0.06 & 33 & 103 \\
{\tt A5}  &  0.001  &   1.6$\times$10$^{-3}$  &  4 & 479  &  0.004  &  0.04 & 100 & 168 \\
{\tt A6}  &  0.001  &   4.3$\times$10$^{-4}$  &  4 & 1506  &  0.004  &  0.04 & 333 & 179 \\\hline
{\tt B1}  &  0.01  &   1.3$\times$10$^{-2}$  &  67 & 67  &  0.054  &  0.53 & 1 & 11 \\
{\tt B2}  &  0.01  &   7.7$\times$10$^{-3}$  &  43 & 144  &  0.035  &  0.34 & 3 & 18 \\
{\tt B3}  &  0.01  &   4.4$\times$10$^{-3}$  &  40 & 406  &  0.032  &  0.32 & 10 & 19 \\
{\tt B4}  &  0.01  &   2.5$\times$10$^{-3}$  &  40 & 1333  &  0.032  &  0.31 & 33 & 20 \\
{\tt B5}  &  0.01  &   1.8$\times$10$^{-3}$  &  38 & 3895  &  0.031  &  0.3 & 100 & 20 \\
{\tt B6}  &  0.01  &   2.5$\times$10$^{-3}$  &  36 & 12181  &  0.029  &  0.28 & 333 & 22 \\\hline
{\tt C1}  &  0.05  &   1.7$\times$10$^{-2}$  &  169 & 169  &  0.135  &  1.31 & 1 & 4 \\
{\tt C2}  &  0.05  &   1.2$\times$10$^{-2}$  &  167 & 557  &  0.133  &  1.3 & 3 & 4 \\
{\tt C3}  &  0.05  &   1.2$\times$10$^{-2}$  &  157 & 1572  &  0.125  &  1.22 & 10 & 5 \\
{\tt C4}  &  0.05  &   1.3$\times$10$^{-2}$  &  141 & 4706  &  0.112  &  1.09 & 33 & 5 \\
{\tt C5}  &  0.05  &   9.0$\times$10$^{-3}$  &  125 & 12531  &  0.1  &  0.97 & 100 & 6 \\
{\tt C6}  &  0.05  &   9.3$\times$10$^{-3}$  &  104 & 34837  &  0.083  &  0.81 & 333 & 7 \\
\end{tabular}}\end{table}

\begin{figure*}
\begin{tabular}{ccc}
  \includegraphics[width=0.65\columnwidth]{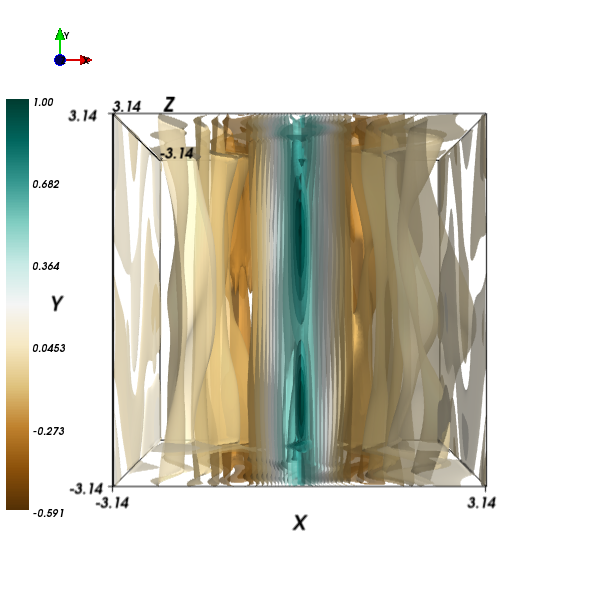}
&  \includegraphics[width=0.65\columnwidth]{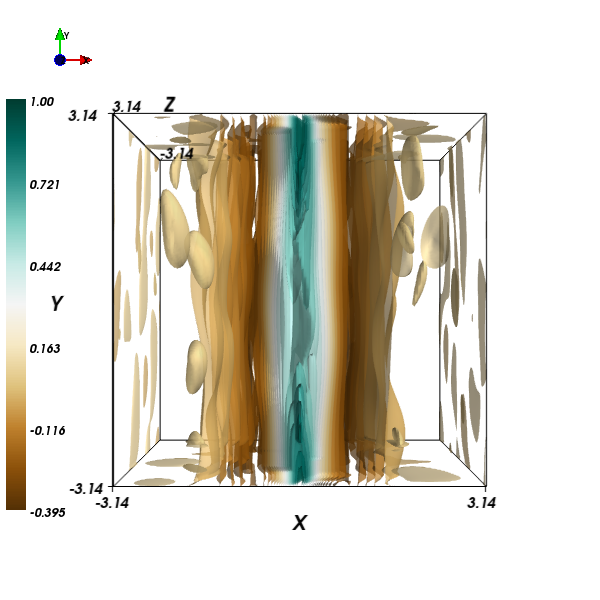}
&  \includegraphics[width=0.65\columnwidth]{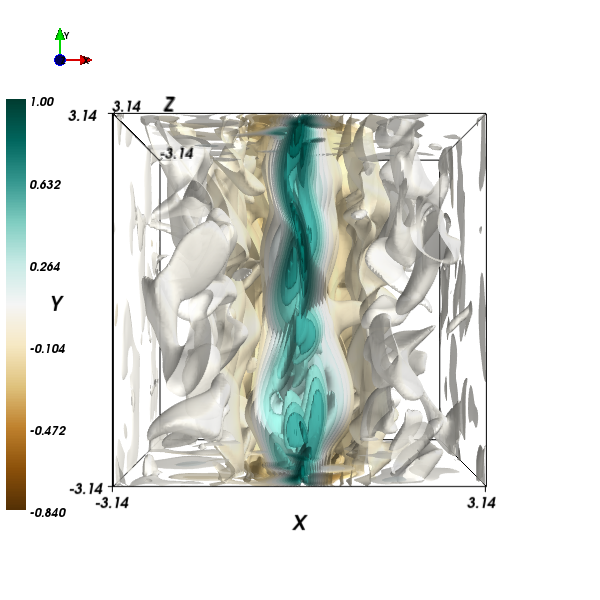}\\
{\tt A2}&{\tt A4}&{\tt A6}\\
  \includegraphics[width=0.65\columnwidth]{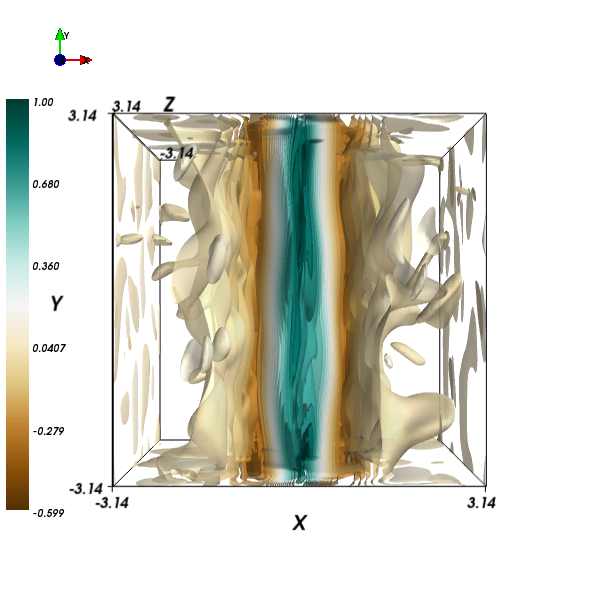}
&  \includegraphics[width=0.65\columnwidth]{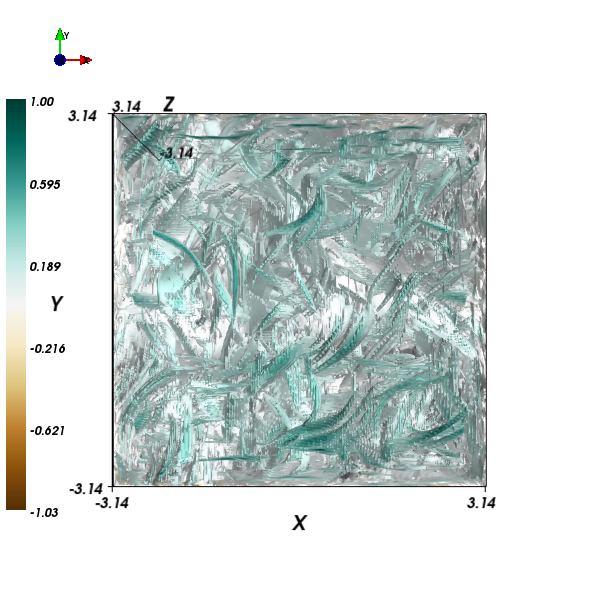}
&  \includegraphics[width=0.65\columnwidth]{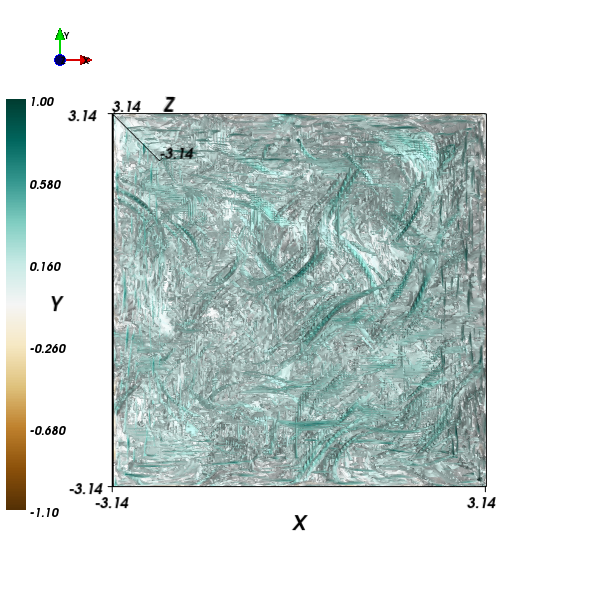}\\
{\tt A5}&{\tt B5}&{\tt C5}
\end{tabular}
\caption{ \label{fig:jz}
Three dimensional pseudocolor plots of $z$ component of the current $\JJ$,   as viewed from XY plane for 
six different runs. Top panels, from left to right : Runs {\tt A2}, {\tt A4}  and {\tt A6}. 
Bottom panels, left to right: Runs {\tt A5}, {\tt B5} and {\tt C5}}
\end{figure*}

\begin{figure*}
\begin{tabular}{cc}
  \includegraphics[width=0.34\linewidth]{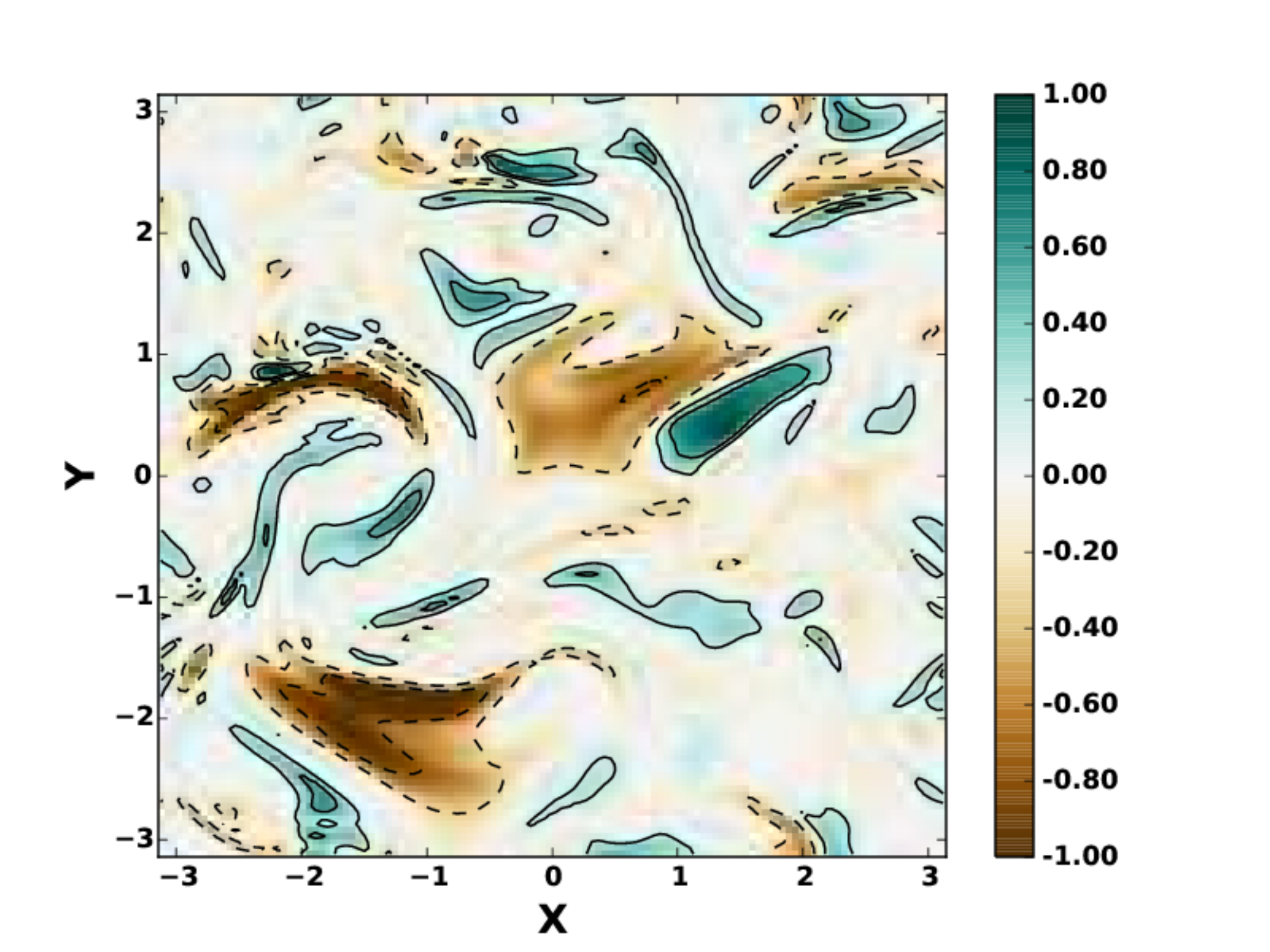} &
  \includegraphics[width=0.34\linewidth]{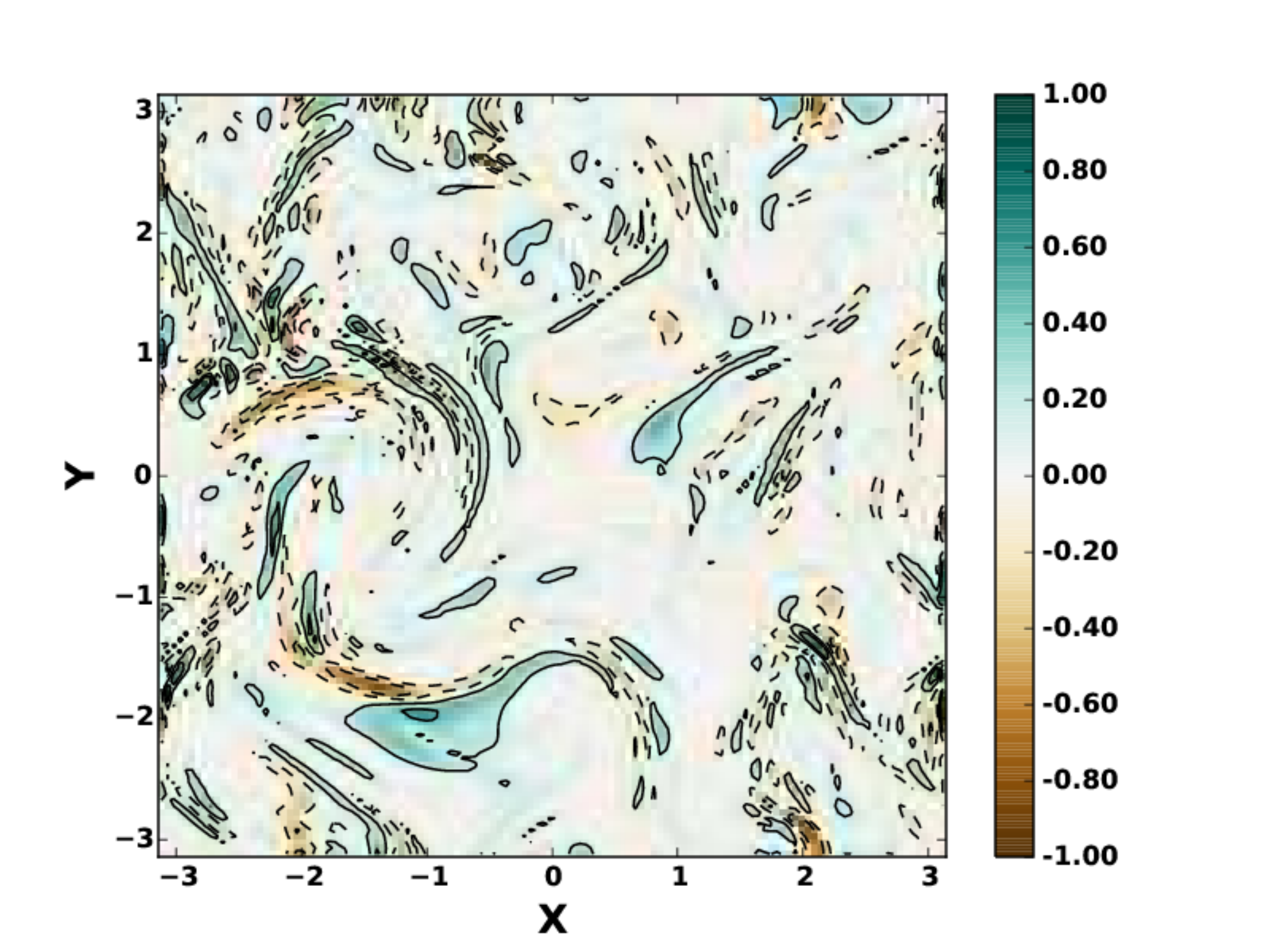} \\
    (a) $B_y(z=L/2)$ & (b) $J_z(z=L/2)$ \\
  \includegraphics[width=0.34\linewidth]{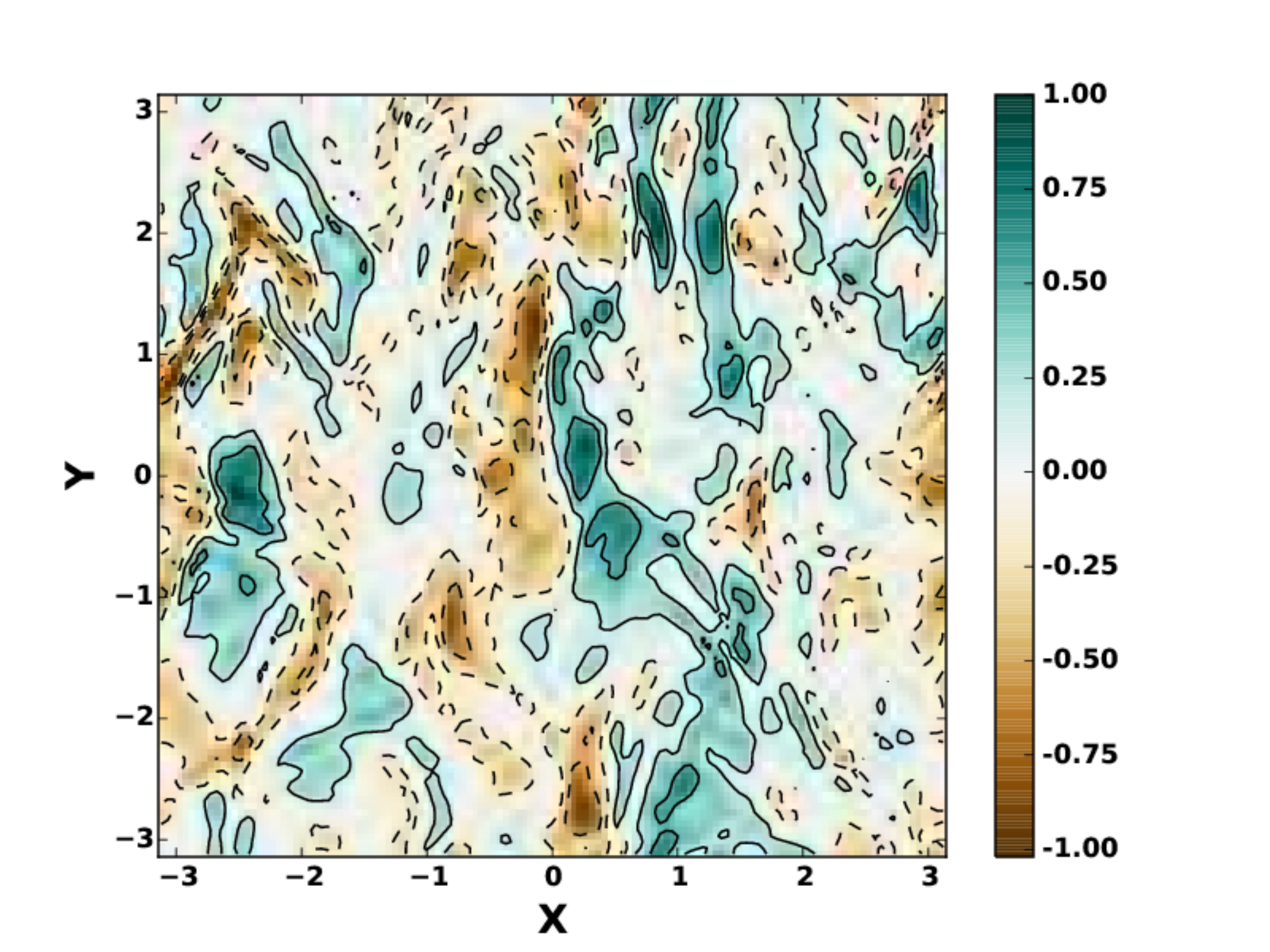} &
  \includegraphics[width=0.34\linewidth]{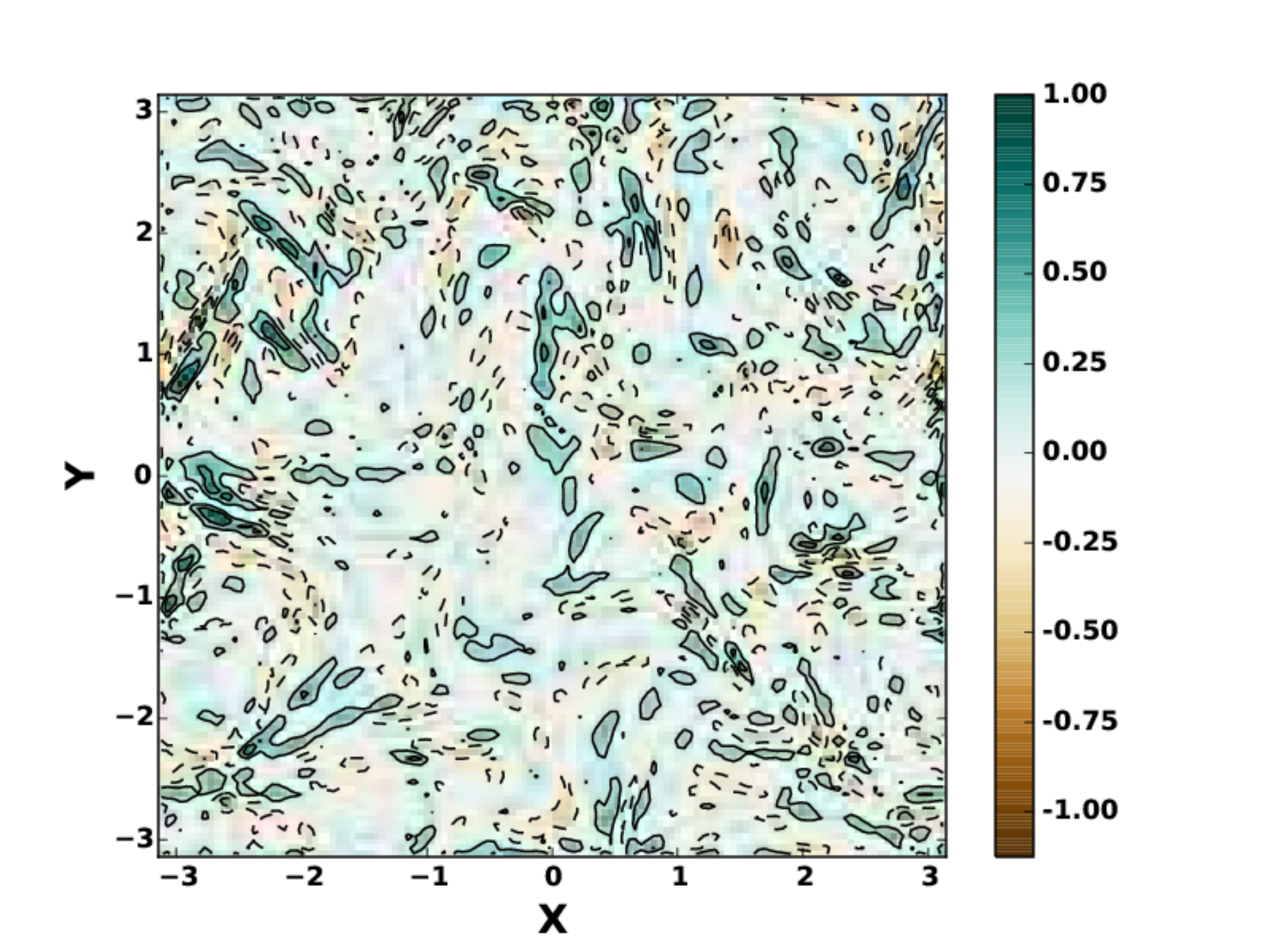} \\
  (c) $\avgz{B_{y}}$ & (d) $\avgz{J_{z}}$
\end{tabular}
\caption{ \label{fig:C4}
Top panels show XY plane snapshot of $B_{y}$ and $J_{z}$ at
$z=L/2$.  The contours are plotted at 25$\%$, 40$\%$, 60$\%$ 
and 75$\%$ levels of the individual peak.
Bottom panels from left to right, 2D pseudocolor plots of z-averaged  
$\avgz{B_y}$ and $\avgz{J_z}$ at a typical time 
during reconnection from the run {\tt C4}. } 
\end{figure*}

\section{Results and Discussion}
\label{sec:results}
We divide the results in two parts. In the first part we discuss our reconnection runs 
without any test particles. The energisation of the test particles is discussed next, 
in the section~\ref{sec:tpart}.

\subsection{Fast Reconnection}
\label{sec:recon}
Our runs can be divided into three families, {\tt A}, {\tt B} and 
{\tt C}, in sequence of increasing intensity of turbulence. 
Within each family, we have performed six runs ranging from e.g., 
{\tt A1} to {\tt A6}, by decreasing the magnetic diffusivity but
keeping other parameters the same. 
The change in $\eta$ gives rise to a decrease in $\Rm$ and a weak
decrease in $\Ma$, consequently the Lundquist number also decreases.  


Classified in this fashion, the run {\tt A1} has the lowest
turbulent intensity and the largest Lundquist number and
the run {\tt C6} the highest turbulent intensity and 
the lowest Lundquist number. 
Table \ref{tab:para} lists various dimensionless quantities for all the runs. 

As mentioned in Section~\ref{inbc}, we start our simulations  
with an initial magnetic field that gives rise to the tearing mode
instability. 
Following \cite{lou+uzd+sch+cow+you09}, we have added a magnetic
forcing term in the induction equation, such that the magnetic
flux consumed by the reconnection is replenished.
In about one large-eddy-turnover-time after the start of our simulation,
we observe the first reconnection event and then we observe the 
magnetic flux building up till the next reconnection event happens.
Eventually, we obtain a statistically stationary state with repeated reconnection
events. 
In \fig{fig:jz} we show the 3D structures of the current sheet by 
plotting a pseudo-color plot of the z-component of the current, $J_{z}$. 
In the top panel of \fig{fig:jz}, we show $J_{z}$ for three runs of family {\tt A}.
We find that as the Lundquist number decreases, the current sheet develops more 
and more small scale structures. 
In the bottom panel of \fig{fig:jz} we show the impact of the
increasing forcing 
amplitude on the current 
sheet structures; numerous small scale structures develop as
the result of increased turbulent intensity.
\cite{sam+lou+uzd+sch+cow09} have suggested that these small scale 
structures are responsible for fast turbulent reconnection.  
In the runs with large amount of turbulent intensity, the effects of 
turbulence can be quite dramatic. To demonstrate this, we show in the
upper panel of \fig{fig:C4}, a contour plot $B_y$ and $J_z$ 
in the $x-y$ plane for $z=L/2$.  The fluctuations due to turbulence
are so large that the current sheet is practically invisible. 
On averaging the same snapshot over the $z$ direction we obtain 
$\avgz{B_y}$ and  $\avgz{J_z}$ which we plot in the bottom panel. 
The averaging decreases the fluctuations and a hint of current sheet structure 
begins to reappear in $\avgz{B_y}$, though it remains hard to discern
in $\avgz{J_z}$. 

At present, there does not exist one unique prescription to measure
the magnetic reconnection rate, particularly in the presence of 
turbulence, see e.g., \cite{comisso2016value} and 
references therein for a critical summary.
In two dimensional laminar cases, the reconnection rate is the
rate of change of vertical magnetic flux at the center of the current
sheet. \cite{lou+uzd+sch+cow+you09} have suggested a generalisation of this 
to the turbulent case where they have measured the vertical magnetic flux
at a fixed point in the grid (at the center of the laminar current sheet)
and have fitted a line to the minima of such a time series to estimate
the reconnection rate. \cite{hua+bha10} in 2D simulations, and  
\cite{ber13} in 3D cases, have estimated reconnection rate by
measuring the rate of change of the width of the current sheet. 
\cite{kow+laz+vis+otm09} have measured it by using a contour 
encompassing the reconnection region and calculating the flux that 
flows into the region.   
This incoming flux can also be estimated by finding out the typical
 speed of the inflow towards the current
 sheet~\citep{jab+bra+mit+kle+rog16}. 

\begin{figure}
\begin{center}
\begin{tabular}{c}
\includegraphics[width=0.95\columnwidth]{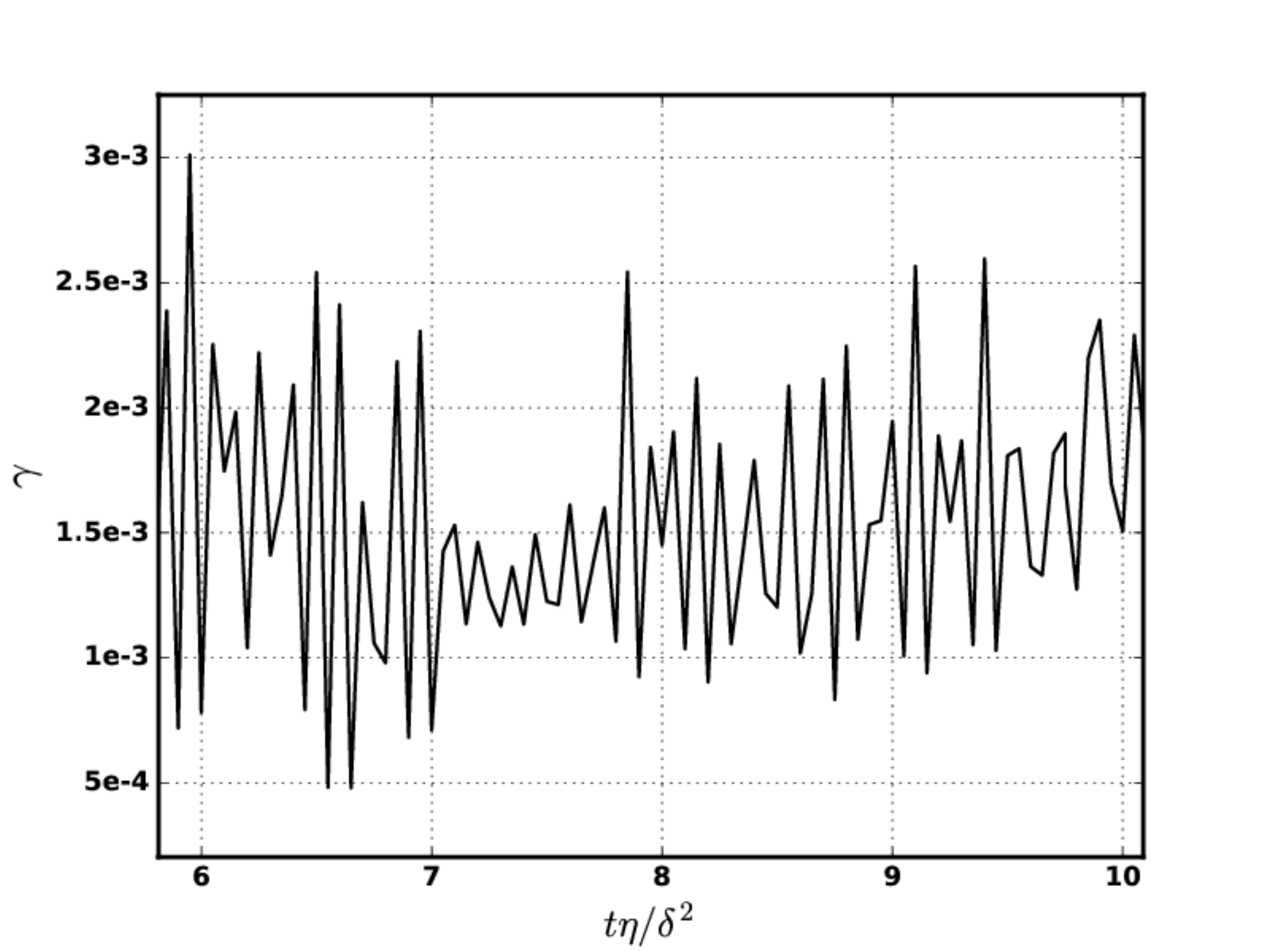}
\end{tabular}
\end{center}
\caption{ A portion of the reconnection rate ($\gamma$) timeseries for {\tt A1} in the steady state. 
 \label{fig:mr_vs_t}}
\end{figure}

In this paper, we suggest a prescription that is similar to the one suggested by 
\cite{kow+laz+vis+otm09}. Consider the midplane ($y=0$) of
our domain (in view of statistical homogeneity along the $y$ direction
 any $x-z$ plane can be considered). In this plane consider 
a closed contour $\mathcal{C}$ which is a rectangle in the x-z plane
whose sides are from  $x=-\delta$ to $x=+\delta$, and $z=-L_z/2$ to
 $z=L_z/2$. The rate-of-change of flux of the $y$ component of the magnetic field
through this contour is given by 
 \begin{eqnarray}
 \delt \Phi &=& -\oint_{\mathcal{C}} \EE \cdot \dl \\
           &=& -\left [\int^{z=-\Lzbytwo}_{z=\Lzbytwo,x=-\delta}\EE\cdot\dl 
                       +\int^{z=\Lzbytwo}_{z=-\Lzbytwo,x=\delta}\EE\cdot\dl
                     \right] \/.
 \end{eqnarray}
 The second equality follows from  the periodic boundary conditions.
 But this flux is always close to zero because it consists of (roughly)
 equal amount of positive and negative flux, in other words the two
 integrals are (roughly) equal and opposite in sign. Hence a
 reasonable estimate of the rate-of-change of flux of only one sign
 can be obtained by reversing the direction of the second line
 integral and dividing by two. More precisely, we suggest the
 following expression:
\begin{equation}
\Vrec = -\frac{1}{2B_0L_z}\left[\int^{z=-\Lzbytwo}_{z=\Lzbytwo,x=-\delta}\EE\cdot\dl 
                       +\int^{z=-\Lzbytwo}_{z=\Lzbytwo,x=\delta}\EE\cdot\dl
                        \right] \/.  
\end{equation}
This has the advantage that the relevant quantities are calculated
away from the wandering current sheet. 
We non-dimensionalise $\Vrec$ by the speed-of-sound, to define,
$\gamma \equiv \Vrec/\cs$, 
which we call the reconnection rate. 
The reconnection rate is a highly fluctuating quantity as a function
of time even for runs that have low turbulent intensity,
as can be seen in the time-series of $\gamma$ shown in
\fig{fig:mr_vs_t}.  
We take the maximum value of $\gamma$
calculated over a large window of time, $\gmax$ as a measure of 
the reconnection rate. The values of $\gmax$ for all our runs
are listed in table~\ref{tab:para}.

\begin{figure}
\begin{center}
\begin{tabular}{c}
\includegraphics[width=0.95\columnwidth]{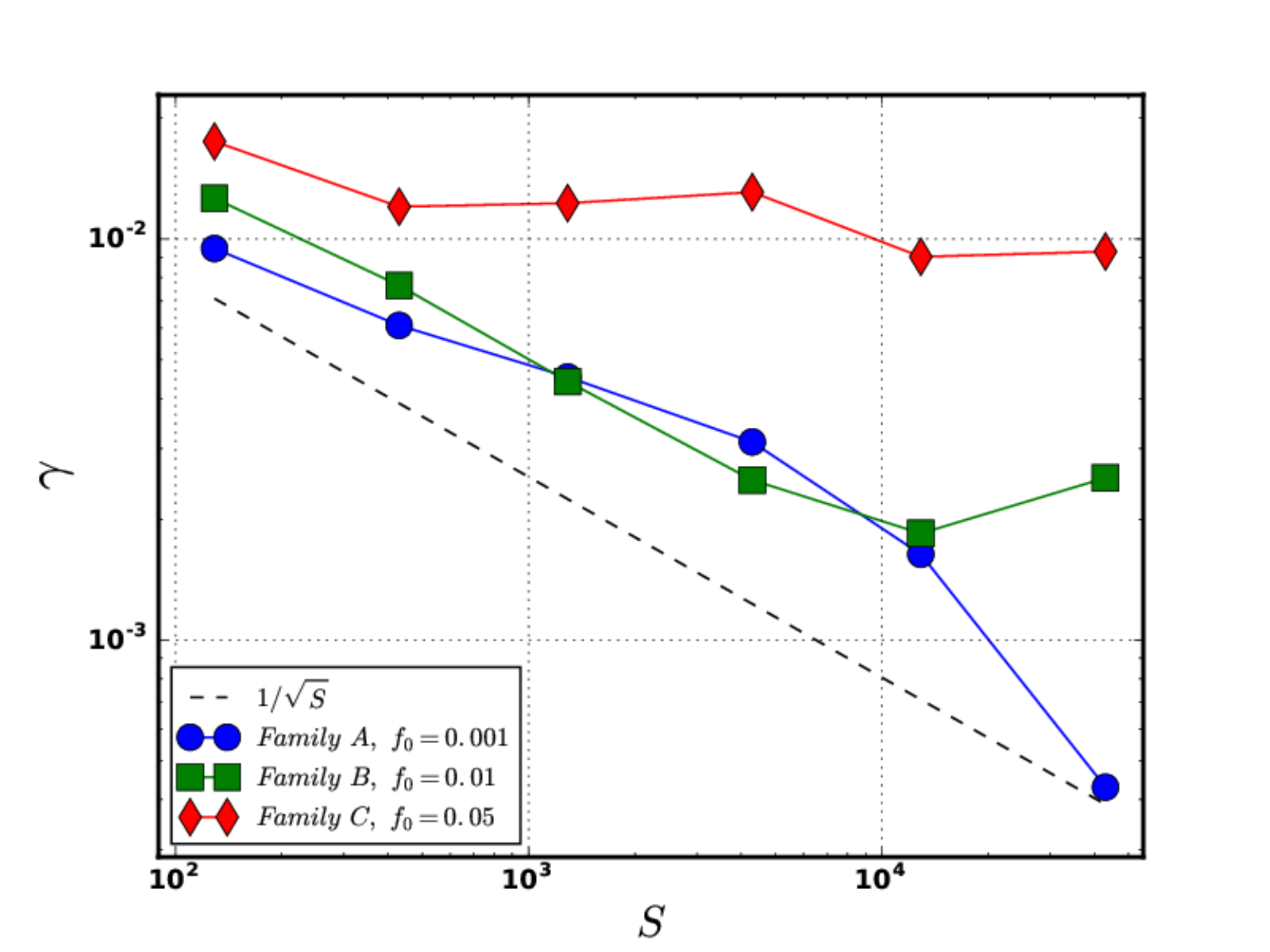}
\end{tabular}
\end{center}
\caption{ The maximum reconnection rate ($\gmax$) versus Lundquist
  number for different turbulent intensities. \label{fig:mr_vs_s}}
\end{figure}

Next, we plot in \fig{fig:mr_vs_s} the maximum reconnection rate
$\gmax$ as a function of the Lundquist number, 
$\S\equiv (\delta \VA)/\eta$, 
We find that the for the runs with the lowest turbulent intensity -- 
the family {\tt A} -- the dependence of $\gmax$ on $\S$ is consistent
which is the Sweet-Parker scaling, which is plotted as a dashed line
in the same figure. If we use the mean value of $\gamma$ over the same
time window instead of its maximum value, we obtain similar results. 
This provides a posteriori justification for using the
prescription for $\gamma$ that we have used.
For the other two families, {\tt B} and {\tt C}, the 
we obtain significant departure from the Sweet-Parker 
behaviour.
For the family {\tt B}, $\gmax$ follows the Sweet-Parker
behaviour upto approximate $\S \approx 10^4$ beyond 
which the dependence on $\S$ is shallower. 
For the family {\tt C} the departure from Sweet-Parker scaling happens
at about $\S \sim 3\times 10^{2}$ beyond which $\gmax$ is almost
a constant as function of $\S$.
Hence the runs in family {\tt C} are clearly in the regime of fast reconnection.
The same behaviour has been observed by   \cite{lou+uzd+sch+cow+you09}
in a similar setup but in 2D. 

\begin{figure}
\begin{center}
\begin{tabular}{c}
\includegraphics[width=0.95\columnwidth]{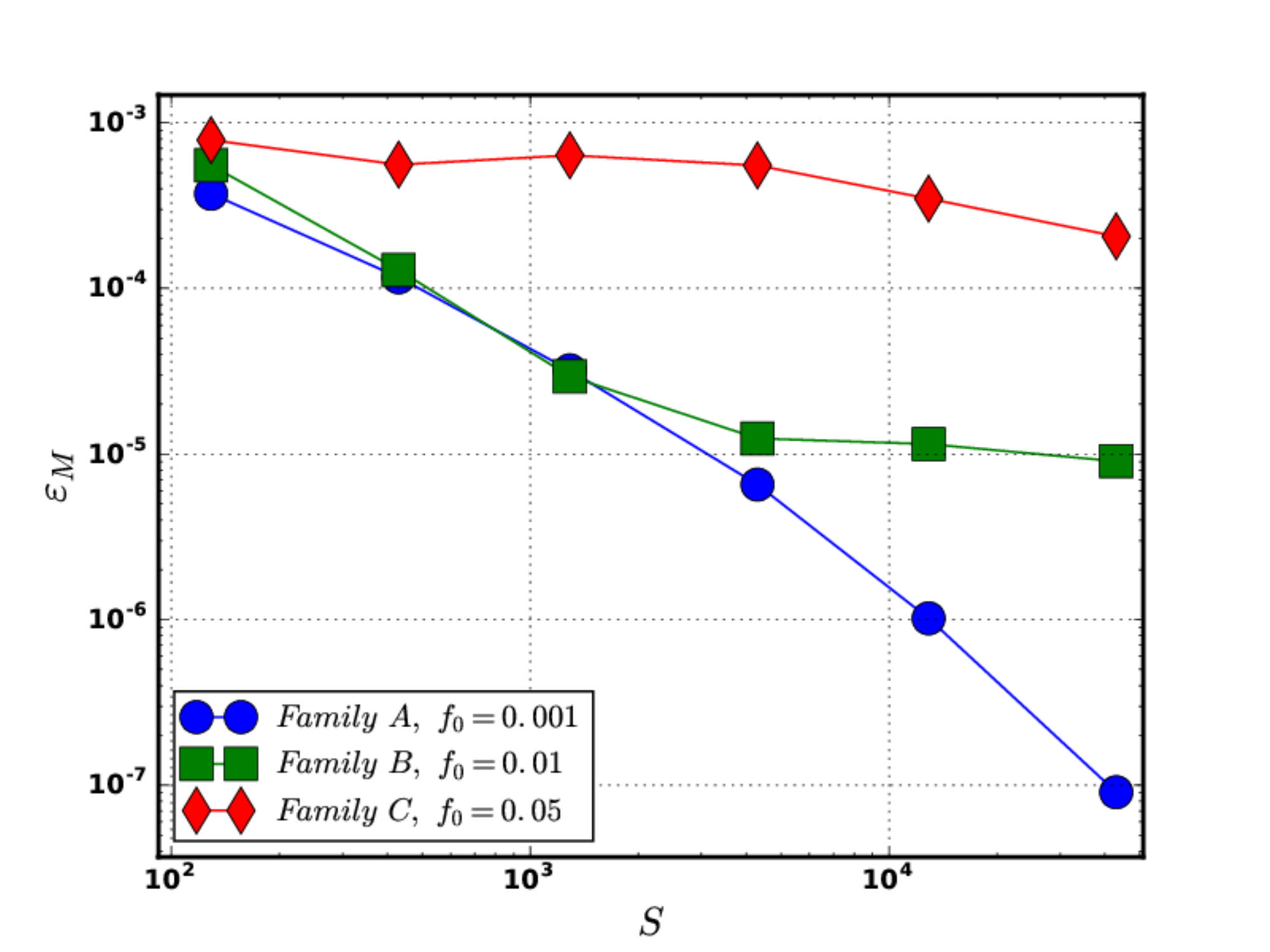}
\end{tabular}
\end{center}
\caption{The rate of dissipation of magnetic energy as a function of the Lundquist number for 
different forcing and diffusivity. Note: Timeseries average of $\epm$ is plotted on the y-axis.
\label{fig:epM_vs_s}}
\end{figure}

Another way to characterise magnetic reconnection is to calculate the
volume-averaged non-dimensional magnetic energy dissipation 
rate~\citep{ois+mac+col+tam15} 
\begin{equation}
\epm \equiv (1/V) \eta \int J^2 dV \/,
\label{eq:epm}
\end{equation}
normalized by $\cs^3/\delta$,
as a function of the Lundquist number $\S$. 
This quantity shows a behaviour qualitatively very similar to the reconnection
rate, for the {\tt A} runs, $\epm$ decreases with $\S$ for large $\S$
but for the {\tt B} and {\tt C} runs -- the runs with high turbulent
intensity -- $\epm$ becomes almost independent of $\S$ for large $\S$;
as shown in \fig{fig:epM_vs_s}. 
The advantage of using $\epm$ as opposed to $\gamma$ as an indicator 
of turbulent reconnection is that the former being a volume averaged
quantity is less noisy in turbulent simulations.  
To summarise, we have established that in our model of
3D tearing-mode reconnection,  fast magnetic reconnection
appears as a result of turbulence. 

\subsection{Energisation of particles}
\label{sec:tpart}
\begin{figure}
\begin{center}
\begin{tabular}{c}
\includegraphics[width=0.95\columnwidth]{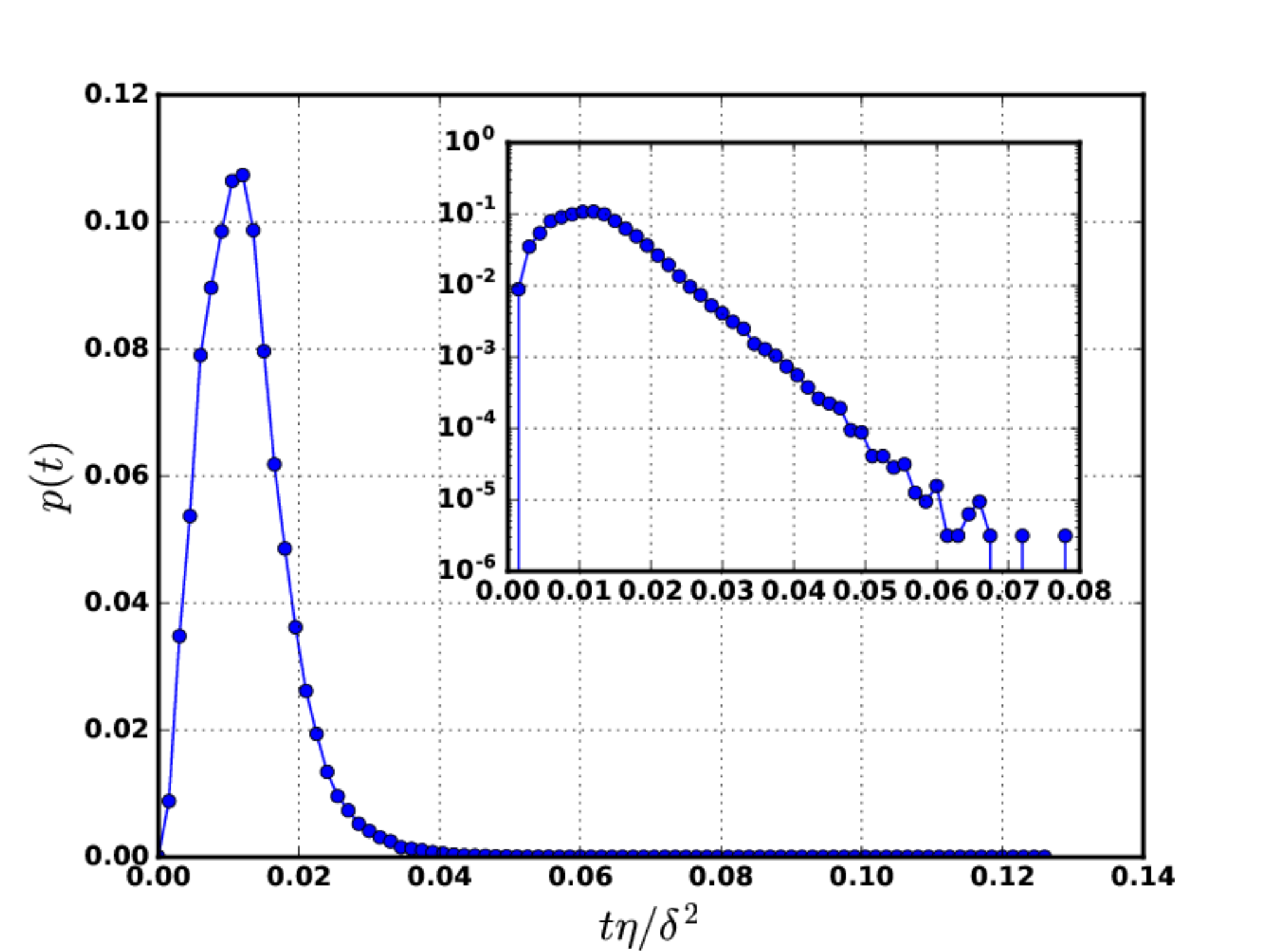}
\end{tabular}
\end{center}
\caption{\label{fig:etime} The fraction of particles that reach the boundary of our
  simulations between time $t$ to $t+dt$ versus time $t$. The origin
  of time is taken to be the instance when the first particle reaches
  the boundary.  }
\end{figure}

\begin{figure}
  \includegraphics[width=0.95\columnwidth]{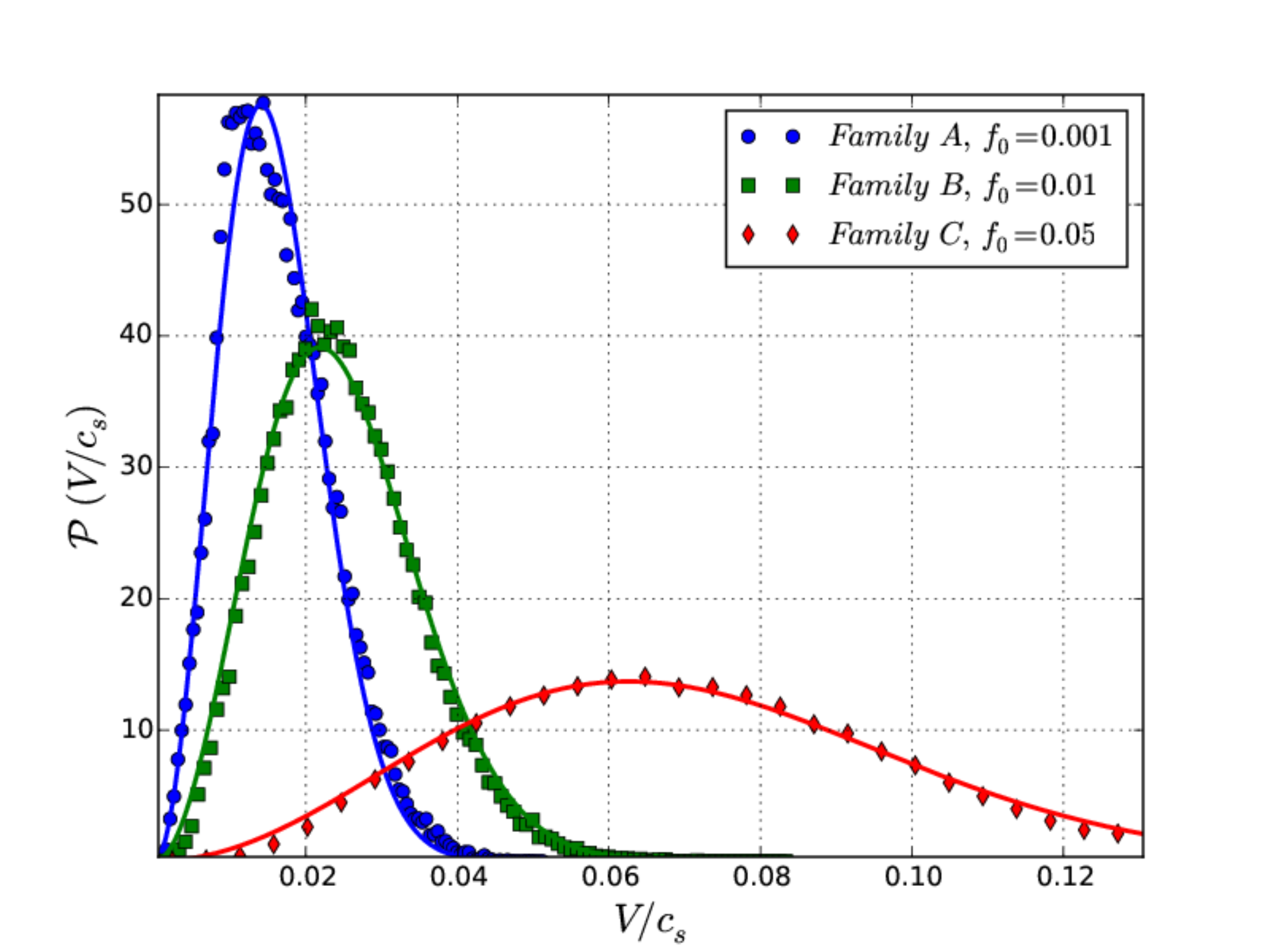}
\caption{ \label{fig:emerg} The speed distribution of the particle removed from the box
for {\tt A4}, {\tt B4} and {\tt C4}. The dashed lines are the
Maxwellian fits to each histogram. 
The $\sigma$ parameter for {\tt A4}, {\tt B4} and {\tt C4} are 0.014, 0.021 and 0.062 respectively.}
\end{figure}

Once a reconnection simulation has reached a statistically stationary state, we
introduce the test-particles with random positions within 
$x=-\delta$ to $x=\delta$, but with zero initial velocity. 
The particles get energised and eventually reach the 
boundary of our simulation domain, where they are removed from the simulation. 
Hence the particles are not allowed to be repeatedly 
energised by the same reconnecting region. 
In \fig{fig:etime} we plot the fraction of particles, $p(t)$, that reach the
boundary of our domain between time $t$ and $t+dt$ as a function of time $t$. 
Curiously, the function $p(t)$ has an exponential tail as shown in 
the inset of \fig{fig:etime}. 
Next, we plot the PDF of the speed of the ejected particles for three 
different representative runs {\tt A4}, {\tt B4} and {\tt C4} in 
\fig{fig:emerg}. All such PDFs are well approximated by a normalised Maxwellian
distribution, 
\begin{equation}
\cP(V) =
\sqrt{\frac{2}{\pi}} \bigg(\frac{V^{2}}{\sigma^{3}} \bigg) \exp\left[-\frac{V^2}{\sigma^2}\right]
\label{eq:max}
\end{equation}
with the variance $\sigma$ a function of the different parameters
of the simulation. To demonstrate the accuracy of this 
fit, we have also plotted in \fig{fig:emerg} the Maxwellian
fits to the PDFs. 
A Maxwellian distribution of speed of particles
energised by the Fermi mechanism has also been observed in
a much simpler context before~\citep{mit+bra+das+nik+ram14}.
Clearly, we do not obtain a non-thermal family of 
energised particles. 
Could the PDF be described by a Maxwellian distribution with a
small non-thermal population at the tail?

\begin{figure}
\begin{center}
\begin{tabular}{c}
\includegraphics[width=0.95\columnwidth]{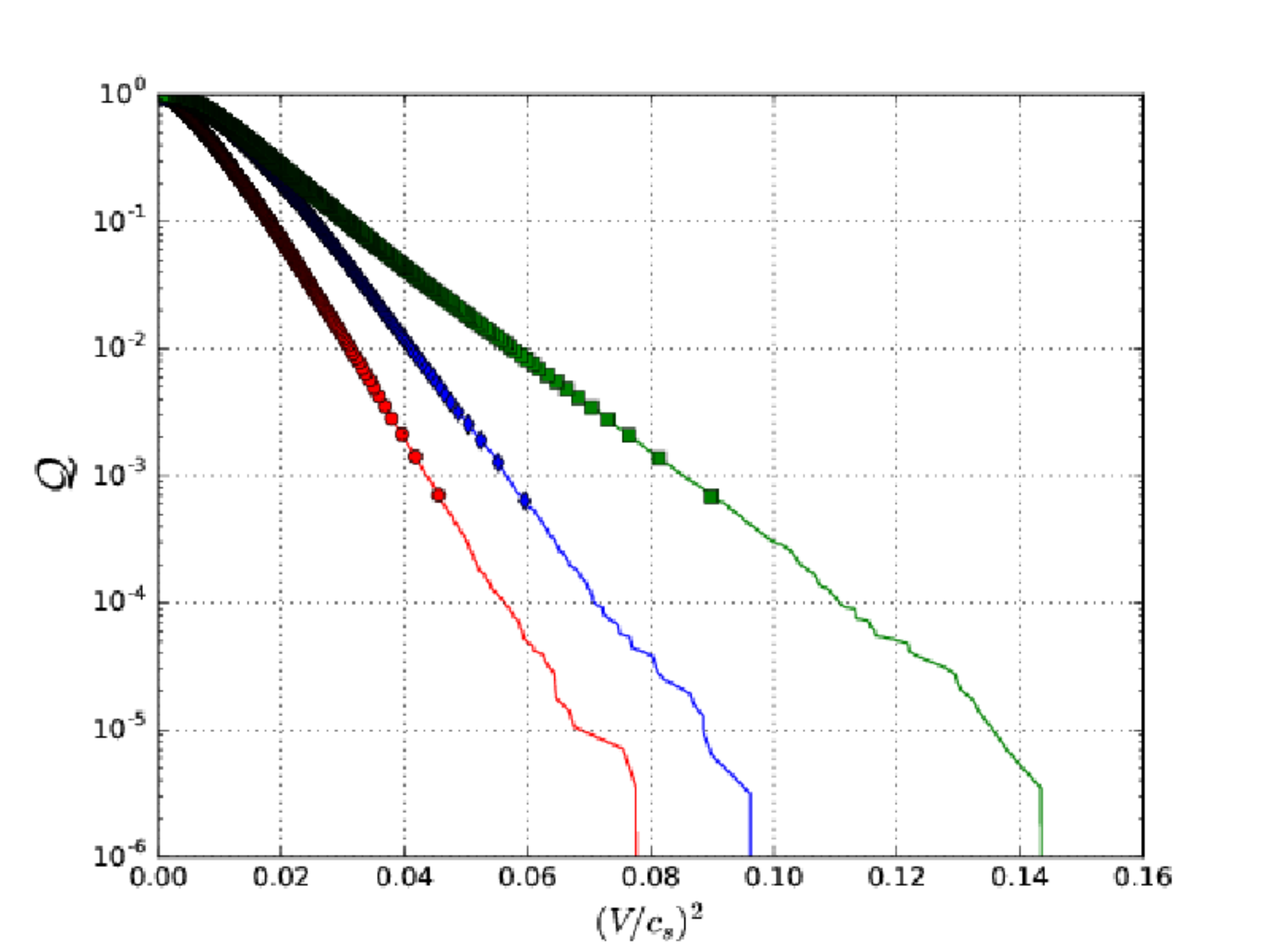}
\end{tabular}
\end{center}
\caption{ \label{fig:cdf} The CDF of kinetic energy of the ejected particles:
 red circles : for  the run {\tt C4}, blue diamonds: for the run {\tt C4a}, which
has the same parameters as the run {\tt C4} except the domain is double in size along
$y$ and $z$ directions, green squares: for the run {\tt C4b}, with same 
parameters as {\tt C4} but with periodic boundary conditions along $y$ and $z$ directions. 
}
\end{figure}

It is generally difficult to obtain reliable information about the
tail of a PDF by plotting histograms because of possible binning
errors. 
Hence instead of studying the tail of the PDF, we have calculated the 
cumulative PDF (CDF) of the kinetic energy of the
particles using rank-order method~\citep{mit+bec+pan+fri05} which allows the
CDF to be free from binning errors 
\footnote{
 By definition, the CDF, $\cQ(X) \equiv \int_{0}^{X}\cP(x)dx$.
To calculate the CDF of a set of data with $N$ samples using rank-order method, 
sort the data in {\it decreasing} order. Assign the maximum value rank $1$, the next 
value Rank $2$ and so on. The quantity we plot in the vertical axis of \fig{fig:cdf}
is this rank divided by the sample size $N$. Clearly, this is equal to $1-\cQ(X)$.
The method is best suited to study tails of CDF. If one is interested in the 
behaviour of the CDF for small values of its arguments, it would be necessary
to sort in {\it increasing} order and then apply the same technique.
}.  
The CDF, for a representative run ({\tt C4}), plotted in \fig{fig:cdf} with red circles,
shows an exponential tail.
An exponential tail of the CDF of kinetic energy implies that the 
PDF also possesses an exponential tail, which in turn implies that 
the tail of the PDF of velocities must be Gaussian.


\begin{figure}
\begin{center}
\begin{tabular}{c}
\includegraphics[width=0.95\columnwidth]{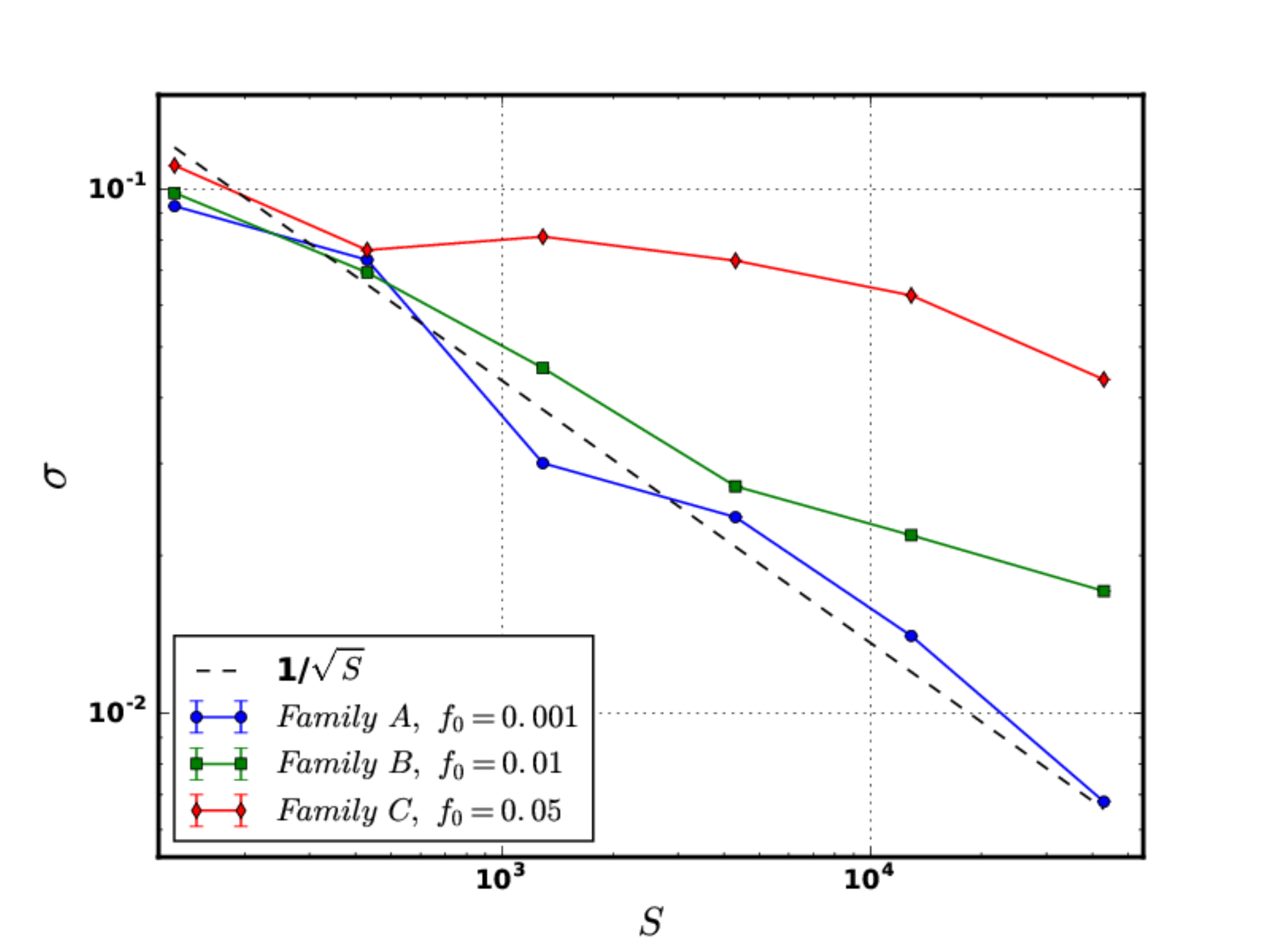}
\end{tabular}
\end{center}
\caption{ The variance of the velocity of the ejected particles as a function of Lundquist number for 
different runs. The dashed line shows $1/\sqrt{\S}$ scaling. Error bars were computed using the deviation obtained from many statistical ensembles of $\sigma$ and are of order of $\sim 0.1\%$.
\label{fig:sigma}}
\end{figure}

As we have now established that the PDF of the energies of the excited
particles is a Maxwellian, it is determined by only one
parameter, $\sigma$. 
The systematic dependence of the variance, $\sigma$, as
a function of $\S$ for the three different families {\tt A}
{\tt B} and {\tt C} are plotted in \fig{fig:sigma}. 
For the {\tt A} family, for which the reconnection rate
follows Sweet-Parker scaling we find the 
$\sigma \sim 1/\sqrt{\S}$ for large $\S$, i.e., the 
typical velocity-scale of the energised particles 
follows the same scaling as that of the reconnection velocity. 
For the other two families, {\tt B} and {\tt C}, the 
we obtain significant departure from the Sweet-Parker 
behaviour.
For the family {\tt B}, $\sigma$ follows the Sweet-Parker
behaviour upto approximate $\S \approx 5\times 10^3$ beyond 
which the dependence on $\S$ is shallower. 
For the family {\tt C} the departure from Sweet-Parker scaling happens
at smaller values of $\S$ and the dependence of $\sigma$ on $\S$ is
a very slow decrease. 

\begin{figure}
\begin{center}
\begin{tabular}{c}
\includegraphics[width=0.95\columnwidth]{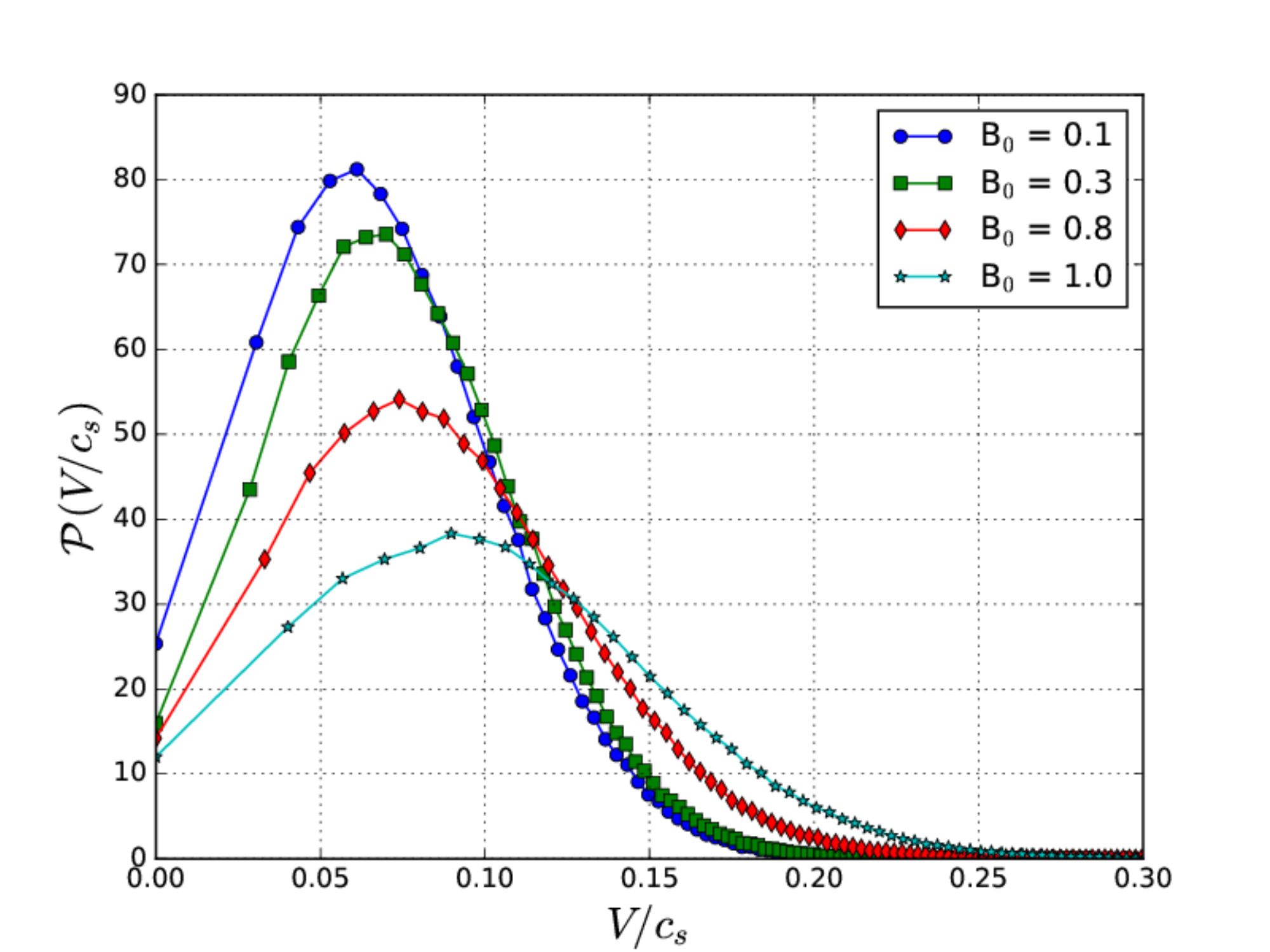}
\end{tabular}
\end{center}
\caption{ \label{fig:PDFB} Probability distribution function of the
  speed of ejected particles for several values of the reconnecting
  magnetic field $B_0 = 0.1,0.3,0.8,1.$, in units of $\cs$. All other
parameters of this run are same as that of {\tt C4} which shows
fast reconnection.
}
\end{figure}

Although the PDF of the speed of the ejected particles is
Maxwellian, the variance, $\sigma$, is not 
determined by the temperature of our simulation, which is
isothermal and same for all the runs. 
The variance, $\sigma$ depends on the Lundquist number and also 
on the magnitude of the reconnecting magnetic field. 
To investigate the latter dependence we run a new set of
simulations. We first select the
run {\tt C4} which is one of our fast reconnection runs. Then 
do a series of runs with the same parameters as {\tt C4} but with different
values of the reconnecting magnetic field. In each of these runs,
we obtain a Maxwellian family of particles but the variance 
increases with strength of the reconnecting magnetic field, $B_0$,
as we show in \fig{fig:PDFB}. 

\begin{figure*}
\begin{tabular}{ccc}
  \includegraphics[width=0.34\linewidth]{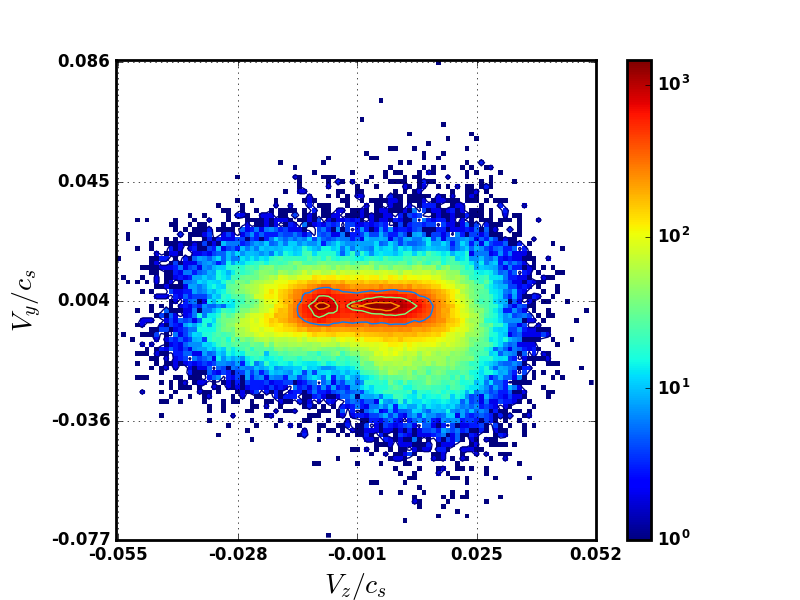} &
  \includegraphics[width=0.34\linewidth]{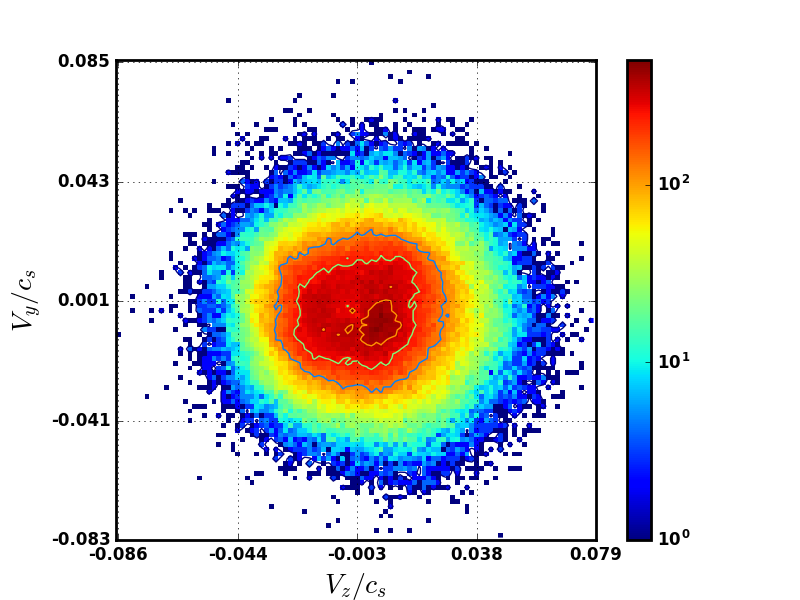} &
  \includegraphics[width=0.34\linewidth]{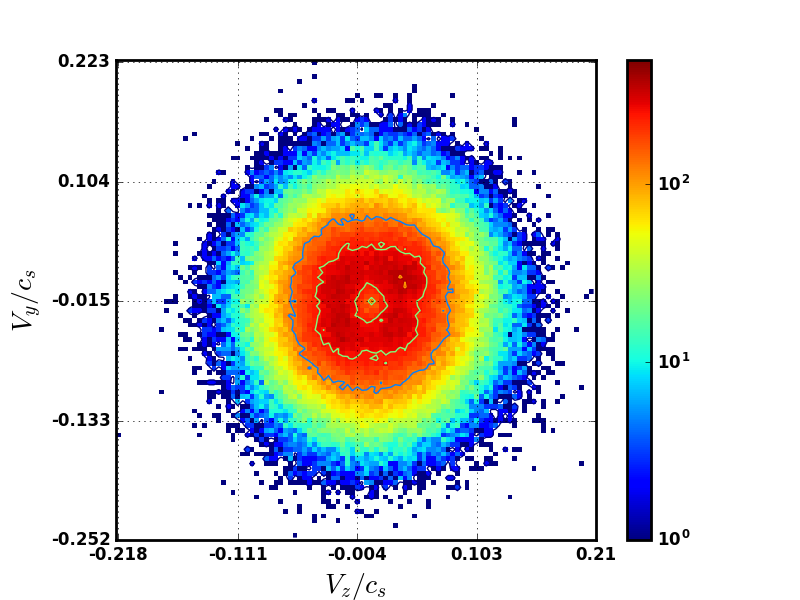}\\
  (a) A4 &(b) B4 &(c) C4\\   
\end{tabular}
\caption{ \label{fig:Vy_Vz}
From left to right, 2D pseudocolor plot of the normalised joint PDFs of $V_{y}$ and 
$V_{z}$ of the ejected particles. 
The contours are plotted at 75$\%$, 50$\%$ 
and 25$\%$ levels of 
the individual peak.}
\end{figure*}

We would also like to point out what may seem to be an inconsistency 
in our results. The setup is clearly anisotropic, with the large-scale 
magnetic field pointing along the $y$ direction and is a function of the
$x$ direction, but the Maxwellian distribution of the speed of the
ejected particles suggests an isotropic distribution of velocities. 
To investigate this point further we have plotted in \fig{fig:Vy_Vz}
a pseudocolor plot of the joint PDF of $V_y$ and $V_z$ 
of the ejected particles for three representative cases; runs 
{\tt A4}, {\tt B4}, and {\tt C4}. 
The first one clearly shows anisotropic behaviour but as the 
strength of the turbulence, which is homogeneous and isotropic, 
increases the joint PDF becomes isotropic too. 
The other two joint PDFs ($V_x$--$V_z$ and $V_x$--$V_y$) show
a similar trend. Simulations of reconnection using particle-in-cell~(PIC)
codes have also observed similar behaviour of the joint PDF 
measured at points away from the reconnection region, although 
near the reconnection region the joint PDF has been shown to be 
far from isotropic~\footnote{Personal communication P. Bourdin} 


The most significant qualitative result of our simulations is 
that the PDF of the charged particles show Maxwellian distribution of 
velocities, or equivalently the tail of the PDF of energy is
exponential. 
Earlier simulations using relativistic PIC codes ~\cite[see e.g.,][figs. 4,5]{oka+pha+kru+fuj+shi10}
have obtained the same result as ours for large plasma beta, but not
for small plasma beta, where magnetic energy dominates over the thermal
energy. Instead ~\cite{dra+swi+che+sha06}  have found power-law tails
in the PDF of energy.
More recent similar simulations, 
e.g., ~\cite{sir+spi14, guo+li+dau+liu14,guo+liu+dau+li15},
have consistently found that the PDF of 
energy has a Maxwellian core superimposed with  power-law tail with an
exponent in the neighbourhood of  $-1$. 
But such power-laws are found in ``parameter regimes where the energy density in the
reconnecting field exceeds the rest mass energy
density''~\citep{guo+li+dau+liu14}.
Presumably, such high magnetic fields are not relevant for the solar
corona. 
An understanding of the power-law tail typically requires
exponential-in-time acceleration of the charged 
particles~\cite[see e.g.][]{dra+swi+che+sha06}
which is in 
in accord with the seminal work of Fermi~\citep{fer49}.
As a counterargument, assume that the electric field, as seen by a charged
particle, in a turbulent medium, has a very short correlation time
and hence can be modelled as white noise.
Then the solution to the problem is a random-walk 
in momentum-space which implies that energy (square of momentum)
grows linearly with time and the PDF of speed would be Maxwellian.
 Both of these have been observed
in numerical simulations of energization in time-periodic 
chaotic magnetic fields~\citep{mit+bra+das+nik+ram14}. 
There are two crucial differences between our work and the
above mentioned PIC simulations: (a) our flows are 
non-relativistic, (b) we use the test-particle approximation.

Simulations, where the test-particles are assumed to be relativistic
although the flow obeys non-relativistic MHD equations,
which further uses the quasi-stationary approximation, claim to have obtained
exponential-in-time acceleration~\citep{kowal2012particle} and  
power-law~\citep{del+dal+kow16} tail in the PDF of energy. 
Could it be that the periodic boundary conditions used in these 
simulations allow the charged particles
to be repeatedly energised by the same reconnection regions and hence
allows for exponential-in-time energisation and also a power-law spectrum?
Furthermore, periodic boundary condition effectively makes the current
sheet infinitely long and allows the possibility of the test-particle
getting energised by repeated collision within the current sheet. 
To test this hypothesis, we have run two more simulations
both with all the parameters same as our run {\tt C4} but the
following differences:
({\tt C4a}) with a box that is double in size along the directions 
parallel to the current sheet ($y$ and $z$ direction);
and 
({\tt C4b}) with periodic boundary conditions. 
The tail of the CDF of energy obtained using rank-order method
for these two simulations are also plotted in \fig{fig:cdf}. 
In both the cases we obtain the same qualitative result; the CDF has 
exponential tail, although quantitatively speaking 
the CDF of run {\tt C4} falls-off the fastest, followed by 
the case ({\tt C4a}) above, followed by the case ({\tt C4b}).

We conclude this section by pointing out the limitations of our work. We have
used test-particles driven by MHD equations. The only way to go beyond
test-particle approximation is to use particle-in-cell methods which have
their own limitations. Furthermore, we have not used the
oft-used quasi-stationary approximation but at a price. If we consider
actual electrons or ion (e.g., proton) then in the corona the estimated
Lorentz number would be of the order of a $10^{16}$, in other words
the typical gyrofrequency of the electron is approximately $10^{16}$
times the typical frequency (one over the time scale of largest
eddies of turbulence) of the largest eddies of turbulence. It is
impossible, even with the present state-of-the-art computing, to
resolve such a wide range of time scales. So we have chosen to use the
smallest Lorentz number that we could use which implies that our
charged particles are effectively very heavy ions. 
This approximation is similar to the arbitrarily
chosen charge-to-mass ratio of electrons in particle-in-cell codes.
Another way out could be to use the
guiding-center approximation for the test-particles. 

\section{Summary}
\label{sec:disc}
To summarise, we have studied two problems in this paper. First we
have studied reconnection in tearing-mode setup in three dimensions 
to find that in the absence of turbulence, reconnection follows
Sweet-Parker scaling but for large enough turbulence it is possible to obtain fast
reconnection.
The topology of magnetic fields change in the presence of turbulence,
developing more  small scale structures. The current sheet fragments
into plasmoids for both large
turbulent intensities and for large Lundquist number. 
The density and characteristic length-scale 
of these islands depend more on the amplitude of the forcing (Fig. \ref{fig:C4})
than the Lundquist number. The fragmentation of the current sheet facilitates the 
formation of numerous reconnection sites and thus can
make fast reconnection possible.
Compared to earlier 2D simulations in a similar setup~\citep{lou+uzd+sch+cow+you09} 
the increase in reconnection rate due to turbulence is more pronounced
in 3D than 2D.
We also show that volume averaged magnetic energy
dissipation rate can act as an useful proxy for the magnetic reconnection rate.  

Next we have studied the energisation of test particles
in this setup. We find that the test particles are indeed energised
but the PDF of their speed can be described well by a Maxwellian
distribution, hence we do not find a non-thermal population of
particles. 
We also find that the PDF of the energised particles does
not depend on temperature but rather on the strength of the
reconnecting magnetic field. From this we can conjecture that if the
strength of the reconnecting magnetic field is large enough -- for
example if the magnetic energy is significantly larger than the local 
thermal energy -- the energised particles would be energised to
energies larger than local thermal energies, although their PDF would still
obey a Maxwellian distribution. Such a family of charged particles
could be called suprathermal. 

\section*{Acknowledgements}

We thank the referee for constructive and pertinent suggestions. DM thanks Philippe Bourdin for useful discussion. We have use the software matplotlib to generate the figures in this paper \citep{matplotlib}.
After our paper was posted on the arxiv we obtained useful
feedback which has been taken into account in this version. 
We would particularly like to thank Elisabete M. de Gouveia Dal Pino
and Fan Guo.  
This work was supported in part by the Swedish Research Council Grant
No. 638-2013-9243 (DM).
We acknowledge the allocation of computing resources provided by the
Swedish National Allocations Committee at the Center for Parallel 
Computers at the Royal Institute of Technology in Stockholm. 
A significant part of the work was done during DM's stay in NCRA
and during visits of RS and DO to Nordita. 
The financial support for these visits from NCRA and Nordita 
are gratefully acknowledged. 

\bibliographystyle{mn2e}
\bibliography{sunref}

\end{document}